\documentclass[lettersize,journal]{IEEEtran}
\IEEEoverridecommandlockouts

\usepackage{amsmath,amsfonts}
\usepackage{algorithmic}
\usepackage{array}
\usepackage[caption=false,font=normalsize,labelfont=sf,textfont=sf]{subfig}
\usepackage{textcomp}
\usepackage{stfloats}
\usepackage{url}
\usepackage{verbatim}
\usepackage{graphicx}
\def\BibTeX{{\rm B\kern-.05em{\sc i\kern-.025em b}\kern-.08em
    T\kern-.1667em\lower.7ex\hbox{E}\kern-.125emX}}
\usepackage{balance}
\usepackage{caption}
\usepackage{siunitx}
\usepackage{tikz}
\usetikzlibrary{shapes.geometric, arrows.meta, positioning, calc, fit}

\usepackage{algorithm}

\usepackage{mathtools}	% loads amsmath
\usepackage{amssymb}
\usepackage{amsthm}

\usepackage[utf8]{inputenc}

\usepackage{ifthen}

\usepackage{microtype}

\usepackage{caption}

\usepackage{bm}

\usepackage[draft,nomargin,marginclue,inline]{fixme}
\makeatletter
\renewcommand*\FXLayoutMarginClue[3]{%
  \marginpar[%
  \raggedleft\@fxuseface{margin}\textcolor{red}{\ignorespaces $ \Rightarrow $}]{%
    \raggedright\@fxuseface{margin}\textcolor{red}{\ignorespaces $ \Leftarrow $}}}
\makeatother	% change color and displayed text in margin
\usepackage{pgfplotstable}	% loads pgfplots, tikz
\pgfplotsset{
	discard if/.style 2 args={
        x filter/.append code={
            \edef\tempa{\thisrow{#1}}
            \edef\tempb{#2}
            \ifx\tempa\tempb
                
            \fi
        }
    },
    discard if not/.style 2 args={
        x filter/.append code={
            \edef\tempa{\thisrow{#1}}
            \edef\tempb{#2}
            \ifx\tempa\tempb
            \else
                
            \fi
        }
    }
}

\usepackage{cite}

\usepackage[acronym,shortcuts]{glossaries}
\newacronym{cnn}{CNN}{convolutional neural network}
\newacronym{ula}{ULA}{uniform linear array}

\usepackage{cleveref}

\tikzset{algorithm1/.style={mark options={solid},color=TUMBeamerBlue, line width=\lineWidth, mark=square, dashed}}

\pgfplotscreateplotcyclelist{miko}{%
{color=TUMBeamerRed, dash pattern=on 5pt off 1pt on 1pt off 1pt, line width=1.5pt},%
{color=TUMBeamerOrange, dash pattern=on 5pt off 2pt, line width=1.5pt,
mark=o},%
{color=black, dash pattern=on 5pt off 2pt on 3pt off 2pt, line width=1.5pt},%
{color=TUMDarkerBlue, line width=1.5pt},%
{color=black, dotted, line width=1.5pt},%
{color=TUMBeamerGreen, dotted, line width=1.5pt},%
{color=TUMBeamerYellow, dotted, line width=1.5pt},%
{color=TUMBeamerDarkRed, dotted, line width=1.5pt},%
{color=TUMBeamerLightGreen, dotted, line width=1.5pt},%
{color=TUMIvory, dotted, line width=1.5pt},%
{color=TUMLightBlue, dotted, line width=1.5pt},%
}

\DeclareMathOperator*{\argmax}{arg\,max}

\DeclareMathOperator{\diag}{diag}

\DeclareMathOperator{\expec}{E}

\newcommand{\calN}{\mathcal{N}}
\newcommand{\calO}{\mathcal{O}}

\newcommand*{\C}{\mathbb{C}}

\newcommand*{\R}{\mathbb{R}}

\newcommand{\herm}{{\operatorname{H}}}
\newcommand{\tp}{{\operatorname{T}}}

\definecolor{myblue}{RGB}{30, 100, 200}

\newlength{\leftstackrelawd}
\newlength{\leftstackrelbwd}
\def\leftstackrel#1#2{\settowidth{\leftstackrelawd}%
	{${{}^{#1}}$}\settowidth{\leftstackrelbwd}{$#2$}%
	\addtolength{\leftstackrelawd}{-\leftstackrelbwd}%
	\leavevmode\ifthenelse{\lengthtest{\leftstackrelawd>0pt}}%
	{\kern-.5\leftstackrelawd}{}\mathrel{\mathop{#2}\limits^{#1}}}

\newcommand{\mbC}{\bm{C}}

\newcommand{\mbF}{\bm{F}}

\newcommand{\mbK}{\bm{K}}

\newcommand{\mbP}{\bm{P}}
\newcommand{\mbQ}{\bm{Q}}

\newcommand{\mbS}{\bm{S}}
\newcommand{\mbT}{\bm{T}}
\newcommand{\mbU}{\bm{U}}
\newcommand{\mbV}{\bm{V}}

\newcommand{\mba}{\bm{a}}

\newcommand{\mbc}{\bm{c}}

\newcommand{\mbf}{\bm{f}}

\newcommand{\mbh}{\bm{h}}

\newcommand{\mbk}{\bm{k}}

\newcommand{\mbn}{\bm{n}}

\newcommand{\mbp}{\bm{p}}
\newcommand{\mbq}{\bm{q}}

\newcommand{\mbu}{\bm{u}}
\newcommand{\mbv}{\bm{v}}
\newcommand{\mbw}{\bm{w}}

\newcommand{\mby}{\bm{y}}

\newcommand{\mbdelta}{{\bm{\delta}}}

\newcommand{\mbSigma}{{\bm{\Sigma}}}

\newcommand{\mbmu}{{\bm{\mu}}}

\newcommand{\mbzero}{{\bm{0}}}

\newcommand{\hhat}{\hat{\mbh}}

\newcommand{\covhi}{\mbC_i}
\newcommand{\covhk}{\mbC_k}

\usetikzlibrary{intersections,pgfplots.fillbetween,patterns,spy}

\Crefname{figure}{Fig.}{Figs.}
\pgfplotsset{compat=1.15}

\newacronym{AWGN}{AWGN}{additive white Gaussian noise}
\newacronym{BLMMSE}{BLMMSE}{Bussgang LMMSE}
\newacronym{BS}{BS}{base station}
\newacronym{CDF}{CDF}{cumulative distribution function}
\newacronym{CNN}{CNN}{convolutional neural network}
\newacronym{CSI}{CSI}{channel state information}
\newacronym{CSIT}{CSIT}{channel state information at the transmitter}
\newacronym{DFT}{DFT}{Discrete Fourier transform}
\newacronym{DL}{DL}{downlink}
\newacronym{DNN}{DNN}{deep neural network}
\newacronym{DoA}{DoA}{direction of arrival}
\newacronym{EM}{EM}{expectation maximization}
\newacronym{FDD}{FDD}{frequency division duplex}
\newacronym{GMM}{GMM}{Gaussian mixture model}
\newacronym{LMMSE}{LMMSE}{linear minimum mean square error}
\newacronym{LOS}{LOS}{line of sight}
\newacronym{LS}{LS}{least squares}
\newacronym{MSE}{MSE}{mean squared error}
\newacronym{MIMO}{MIMO}{multiple-input multiple-output}
\newacronym{MPC}{MPC}{multi-path component}
\newacronym{MT}{MT}{mobile terminal}
\newacronym{NLOS}{NLOS}{non-line of sight}
\newacronym{NN}{NN}{neural network}
\newacronym{O2I}{O2I}{outdoor-to-indoor}
\newacronym{OMP}{OMP}{orthogonal matching pursuit}
\newacronym{PDF}{PDF}{probability density function}
\newacronym{PGA}{PGA}{projected gradient ascent}
\newacronym{PSD}{PSD}{power spectral density}
\newacronym{SNR}{SNR}{signal-to-noise ratio}
\newacronym{TDD}{TDD}{time division duplex}
\newacronym{UL}{UL}{uplink}
\newacronym{ULA}{ULA}{uniform linear array}
\newacronym{URA}{URA}{uniform rectangular array}
\newacronym{UMa}{UMa}{urban macrocell}
\newacronym{nSE}{nSE}{normalized spectral efficiency}
\newacronym{cCDF}{cCDF}{complementary cumulative distribution function}
\newacronym{MU-MIMO}{MU-MIMO}{multi-user MIMO}
\newacronym{MU-MISO}{MU-MISO}{multi-user MISO}
\newacronym{BD}{BD}{block diagonalization}
\newacronym{RBD}{RBD}{regularized block diagonalization}
\newacronym{RCI}{RCI}{regularized channel inversion}
\newacronym{WMMSE}{WMMSE}{weighted minimum mean square error}
\newacronym{IWMMSE}{IWMMSE}{iterative weighted minimum mean square error}
\newacronym{SWMMSE}{SWMMSE}{stochastic WMMSE}
\newacronym{SVD}{SVD}{singular value decomposition}
\newacronym{SR}{SR}{sum-rate}
\newacronym{CME}{CME}{conditional mean estimator}
\newacronym{ML}{ML}{machine learning}
\newacronym{FLOPS}{FLOPS}{floating-point operations}
\newacronym{OFDM}{OFDM}{orthogonal frequency-division multiplexing}
\newacronym{LTE}{LTE}{Long Term Evolution}
\newacronym{GPS}{GPS}{Global Positioning System}
\newacronym{UMi}{UMi}{urban microcell}
\newacronym{VQ-VAE}{VQ-VAE}{vector quantized-variational autoencoder}
\newacronym{AE}{AE}{autoencoder}
\newacronym{3GPP}{3GPP}{3rd Generation Partnership Project}

\newacronym{MAP}{MAP}{maximum a posteriori}
\newacronym{GNN}{GNN}{graph neural network}

\newcommand{\Nv}{N_{\mathrm{v}}}
\newcommand{\Nh}{N_{\mathrm{h}}}

\begin{document}

\title{Statistical Precoder Design in Multi-User Systems via Graph Neural Networks and Generative Modeling}

\author{Nurettin~Turan,~\IEEEmembership{Graduate Student Member,~IEEE,} Srikar~Allaparapu, Donia~Ben~Amor,~\IEEEmembership{Graduate Student Member,~IEEE,}
Benedikt~Böck,~\IEEEmembership{Graduate Student Member,~IEEE,} Michael~Joham,~\IEEEmembership{Member,~IEEE,} and Wolfgang~Utschick,~\IEEEmembership{Fellow,~IEEE}

\thanks{The authors are with the TUM School of Computation, Information and Technology, Technische Universität München, 80333 Munich, Germany. \textit{Corresponding author: Nurettin Turan} (e-mail: nurettin.turan@tum.de).\\
The authors acknowledge the financial support by the Federal Ministry of
Education and Research of Germany in the program of ``Souver\"an. Digital.
Vernetzt.''. Joint project 6G-life, project identification number: 16KISK002.
}
}

%\markboth{Journal of \LaTeX\ Class Files,~Vol.~18, No.~9, September~2020}%
%{How to Use the IEEEtran \LaTeX \ Templates}

\maketitle

\begin{abstract}

This letter proposes a \ac{GNN}-based framework for statistical precoder design that leverages model-based insights to compactly represent statistical knowledge, resulting in efficient, lightweight architectures.
The framework also supports approximate statistical information in \ac{FDD} systems obtained through a \ac{GMM}-based limited feedback scheme in massive \ac{MIMO} systems with low pilot overhead.
Simulations using a spatial channel model and measurement data demonstrate the effectiveness of the proposed framework.
It outperforms baseline methods, including stochastic iterative algorithms and \ac{DFT} codebook-based approaches, particularly in low pilot overhead systems.

\end{abstract}


\section{Introduction}

\vspace{-1em}
\begin{tikzpicture}[remember picture,overlay]
\node[anchor=south,yshift=14pt,xshift=-134pt] at (current page.south) {{\parbox{\dimexpr\columnwidth-\fboxsep-\fboxrule\relax}{\footnotesize \quad \copyright This work has been submitted to the IEEE for possible publication. Copyright may be transferred without notice, after which this version may no longer be accessible.}}};
\end{tikzpicture}%

Maximizing the sum-rate under a transmit power constraint is crucial to increasing the throughput in massive \ac{MIMO} systems.
Many seminal works, e.g., the \ac{IWMMSE} algorithm \cite{ShRaLuHe11}, assume access to perfect \ac{CSI} about each \ac{MT} during the online phase for precoder design.
More practical approaches, such as \cite{RaSaLu16, ShMa17} rely only on statistical knowledge for precoder design at the \ac{BS}, where the \ac{SWMMSE} \cite{RaSaLu16} is the stochastic counterpart of the \ac{IWMMSE}.
However, the iterative nature of the \ac{SWMMSE} comes with a computational burden that may not meet the latency requirements of massive \ac{MIMO} systems with large antenna arrays and multiple \acp{MT}.%served simultaneously.

Recently, \acp{GNN} have been used for sum-rate maximization under perfect \ac{CSI} assumptions in \cite{ZhGuYa22, LiGuYa24}, and for energy efficiency maximization with perfect statistical knowledge \cite{HeLiLuAiDiNi24}.
Building on \cite{LiGuYa24}, the work in \cite{RiBeJoUt24} extends the \ac{GNN} architecture for the imperfect \ac{CSI} case.
The strong performance in all aforementioned works arises from exploiting the invariance that reordering \ac{MT} indices permutes the precoders accordingly without affecting the sum-rate \cite{ZhGuYa22}.
The \ac{GNN} exploits this expert knowledge by construction, acting as an effective form of regularization.
A significant advantage of \ac{GNN}-based approaches is their scalability to different numbers of served \acp{MT} without requiring retraining.

In \ac{FDD} systems, precoder design typically relies on accurate feedback from the \acp{MT}.
The recently proposed \ac{GMM}-based feedback scheme in \cite{TuFeKoJoUt23} provides a low-complexity, flexible approach that supports different numbers of \acp{MT}, \ac{SNR} levels and pilot configurations without re-training, and is also parallelizable.
The \ac{GMM} learns the overall \ac{PDF} of the channels within a \ac{BS} environment in an offline phase and is shared with \acp{MT} to enable online feedback inference, acting as a generative prior.
So far, the \ac{SWMMSE} has been used with this feedback scheme for precoding. 
To our knowledge, this work is the first to leverage a \ac{GNN} framework for statistical precoder design in \ac{FDD} systems using approximate statistical information.
The contributions of this work can be summarized as follows:

\begin{itemize}
    \item We propose a \ac{GNN}-based precoder design framework for sum-rate maximization using only statistical knowledge of each \ac{MT}'s channel.
    Leveraging model-based insights, we compactly represent statistical knowledge, resulting in efficient, lightweight architectures in massive \ac{MIMO} systems.
    Specifically, the network input per \ac{MT} is a vector parametrizing the covariance matrix, which only scales linearly with the number of \ac{BS} antennas.
    \item The framework is extended to handle approximate statistical information by incorporating the recently proposed \ac{GMM}-based feedback approach, which enables efficient limited feedback in \ac{FDD} systems with low pilot overhead.
    This integration enables overall a flexible \ac{GMM}- and \ac{GNN}-based combined precoder design scheme.
    \item Simulation results with both a spatial channel model and real-world measurement data demonstrate the effectiveness of the proposed framework, outperforming baseline methods, including stochastic iterative algorithms and \ac{DFT} codebook-based approaches, especially in systems with low pilot overhead.
\end{itemize}

\section{System Model and Channel Data}
\label{sec:system_channel_model}

\subsection{Data Transmission Phase}

The \ac{DL} of a single-cell multi-user system where a \ac{BS} with $N$ transmit antennas serves $J$ single-antenna \acp{MT} is considered.
Linear precoding is used, where the precoders $\{\mbv_j\}_{j=1}^J$ satisfy the power constraint $\sum_{j=1}^J \|\mbv_j\|_2^2=\rho$. 
The design objective is the sum-rate, expressed as
\begin{equation} 
\label{eq:inst_sumrate}
    R = \sum_{j=1}^J \log_2 \Bigg(1 + \dfrac{\left|\mbh_j^\tp\mbv_j\right|^2}{ \sum_{j^\prime \neq j} \left|\mbh_j^\tp\mbv_{j^\prime}\right|^2 + \sigma_n^2}\Bigg),
\end{equation}
where $\mbh_j\in \C^{N}$ is the channel of \ac{MT} $j$ and $\sigma_n^2$ denotes the noise variance.
The \ac{BS} designs the precoders $\{\mbv_j\}_{j=1}^J$ based on either perfect statistical knowledge or, more realistically, based on each \ac{MT}'s feedback information, encoded by $B$ bits.

\subsection{Pilot Transmission Phase}

Before data transmission, the \ac{BS} broadcasts ${n_\mathrm{p}}$ orthogonal pilots, allowing each \ac{MT} to infer its feedback information.
The received signal $\mby_j \in \C^{{n_\mathrm{p}}} $ at each \ac{MT} is given by
\begin{equation} \label{eq:noisy_obs}
    \mby_j = \mbP \mbh_j + \mbn_j,
\end{equation}
where $\mbn_j \sim \mathcal{N}_\C(\mathbf{0}, \mbSigma)$ represents the \ac{AWGN} with $\mbSigma = \sigma_n^2 \mathbf{I}_{{n_\mathrm{p}}}$.
For a \ac{ULA} at the \ac{BS}, a DFT (sub)matrix is selected as the pilot matrix $\mbP$, while a 2D-DFT (sub)matrix is used for a \ac{URA}, cf. \cite{TuFeKoJoUt23}.
Each column $\mbp_l$ of $\mbP^\tp$ is normalized to meet the power constraint $\|\mbp_l\|_2^2=\rho$ for all $l \in \{1,2, \dots, {n_\mathrm{p}}\}$.
We consider systems with reduced pilot overhead, ${n_\mathrm{p}} < N$, due to their practical relevance \cite{BjLaMa16}.

\subsection{Channel Data}

\subsubsection{Spatial Channel Model} \label{sec:scm_data}

We use the spatial channel model outlined in \cite{NeWiUt18}, where the channels are modeled conditionally Gaussian, i.e., \( \mbh_{j} | \mbdelta_j \sim \calN_\C(\mbzero, \mbC_{\mbdelta_j}) \).
The random vectors \( \mbdelta_j \) comprise uniformly distributed main departure angles of the multi-path propagation clusters between the \ac{BS} and each \ac{MT}.
The \ac{BS} uses a \ac{ULA} with $N$ elements, resulting in different Toeplitz-structured covariance matrices $\{\mbC_{\mbdelta_j}\}_{j=1}^J$ for various multi-user scenarios.
This model enables simulations with perfect statistical knowledge for performance benchmarking.

We construct a data set $\mathcal{H} = \{ \mbh^{(m)} \}_{m=1}^{M}$, where each channel sample $\mbh^{(m)}$ is generated by drawing random angles \( \mbdelta^{(m)} \) and sampling \( \mbh^{(m)} \sim \calN_\C(\mbzero, \mbC_{\mbdelta^{(m)}}) \).
The resulting data set represents a \ac{BS} environment with unknown \ac{PDF} $f_{\mbh}$. 

\subsubsection{Real-World Channel Data} \label{sec:meas_data}

Moreover, measurement data are used to evaluate performance in a real-world \ac{UMi} propagation scenario.
The measurements were collected at the Nokia campus in Stuttgart, Germany, in 2017.
The \ac{BS} installed on a rooftop at $\SI{20}{m}$, comprised a \ac{URA} with $\Nv=4$ vertical and $\Nh=16$ horizontal elements ($N=64$).
Horizontal and vertical spacings were $\lambda/2$ and $\lambda$, respectively, with a carrier frequency of $\SI{2.18}{\giga\hertz}$. 
Please see \cite{HeDeWeKoUt19} for details.

Here, a set $\mathcal{H}$ consisting of $M$ measured channels constitutes the \ac{BS} environment with unknown \ac{PDF} $f_{\mbh}$. 

\section{Proposed Framework}

\subsection{GNN-based Precoder Design given Perfect Statistics} \label{sec:gnn_perfect}

To obtain the precoders $\{\mbv_j\}_{j=1}^J$ solely utilizing statistical knowledge about the channel $\mbh_j$ of each \ac{MT}, the key adaptation of the \ac{GNN} compared to the existing works \cite{LiGuYa24, RiBeJoUt24}, is given at the input.
With perfect statistical knowledge, i.e, assuming the knowledge of $\{\mbC_{\mbdelta_j}\}_{j=1}^J$, model-based insights can be exploited to present the statistical knowledge to the \ac{GNN}. 
Specifically, for a \ac{ULA}, the covariance matrix is Toeplitz-structured \cite{NeWiUt18}, and for a \ac{URA}, it is block-Toeplitz with Toeplitz blocks \cite{NoZoSuLo14}.
In both cases, the covariance matrix per \ac{MT} $j$ is fully parametrized by its first row (or column).
Thus, for each \ac{MT} $j$, the first row of the respective covariance matrix is extracted
\begin{equation}
    \mbc_{j} = \mbC_{\mbdelta_j}[1,:].
\end{equation}
Subsequently, the real and imaginary parts of $\mbc_{j}$ are stacked and passed through a feature extractor $\mbF_\theta$ to obtain the input feature vectors $\mbf_j^{(0)}$ which are relevant for the \ac{GNN} processing.
This has the advantage of a reduced input dimension of $2N$ per \ac{MT}, which consequently does not scale with the covariance matrix dimension of $N^2$ and additionally inherently reduces the computational complexity.
To facilitate scalability, the feature extractor $\mbF_\theta$ consists of a single fully connected layer followed by a PReLU activation function, and parameter sharing is considered such that the same feature extractor $\mbF_\theta$ is used for all \acp{MT}, similar to \cite{RiBeJoUt24}.

\begin{figure}
    \centering
    \includegraphics[scale=0.77]{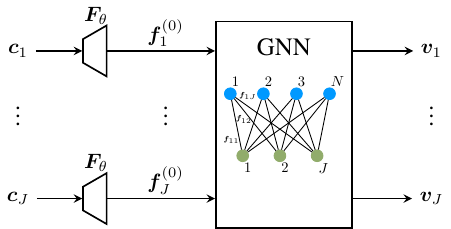}
    \caption{Block diagram of the proposed \ac{GNN}-based precoder design framework.}
    \label{fig:stat_gnn_framework}
    \vspace{-0.5cm}
\end{figure}

As \ac{GNN} architecture, we adopt the Edge-GAT model from \cite{RiBeJoUt24}, originally termed A2D-GNN in \cite{LiGuYa24}.
The \ac{GNN} is structured as a neural network with multiple layers.
Its associated graph is bipartite, where one set of vertices corresponds to the $N$ antennas at the \ac{BS} and another set to the $J$ single-antenna \acp{MT}, and the features are solely associated with the edges of the graph.
For the \ac{GNN} processing, each feature vector $\mbf_j^{(0)} = [\mbf_{1j}^{(0),\tp}, \dots, \mbf_{Nj}^{(0),\tp}]^\tp$ must be at least of dimension $N$ such that it contains all initial features for the edges $(n,j)$ for $n = 1, \dots, N$.
Similar to \cite{RiBeJoUt24}, we set the output dimension of the feature extractor $\mbF_\theta$ such that $\mbf_{nj}^{(0)} \in \mathbb{R}^2$ for $n = 1, \dots, N$ and $j = 1, \dots, J$.
Let $\mbf_{nj}^{(\ell-1)}$ denote the feature vector associated with the edge $(n,j)$ at the input of the $\ell$-th layer, with $\ell = 1,\dots, L$. 
The update rule for the edge $(n,j)$ is then \cite{LiGuYa24, RiBeJoUt24}
\begin{equation}
    \mbf_{nj}^{(\ell)} = f_{\text{act}} ( \mbS^{(\ell)} \mbf_{nj}^{(\ell-1)} + \alpha \sum_{\substack{i=1\\ i \neq n}}^N \mbT^{(\ell)} \mbf_{ij}^{(\ell-1)} + \beta \sum_{\substack{k=1\\k\neq j}}^J \mba^{(\ell)}_{jk} \odot \mbu_{nk}^{(\ell)} )
\end{equation}
with $\mba_{jk}^{(\ell)}$ obtained through a linear attention mechanism as
\begin{align}
    \mbq_{nj}^{(\ell)} = \mbQ^{(\ell)} \mbf_{nj}^{(\ell-1)} \quad \mbk_{nj}^{(\ell)} &= \mbK^{(\ell)} \mbf_{nj}^{(\ell-1)} \quad \mbu_{nj}^{(\ell)} = \mbU^{(\ell)} \mbf_{nj}^{(\ell-1)} \nonumber \\
    \mba_{jk}^{(\ell)} &= \sum\nolimits_{n=1}^N \mbq_{nj}^{(\ell)} \odot \mbk_{nk}^{(\ell)}/N. \label{eq:attention_mech}
\end{align}
The matrices $\mbS^{(\ell)}$, $\mbT^{(\ell)}$, $\mbQ^{(\ell)}$, $\mbK^{(\ell)}$, $\mbU^{(\ell)} \in \mathbb{R}^{{M_\ell} \times {M_{\ell-1}}}$, denote the trainable parameters of layer $\ell$ and are used to update the features associated with all edges $(n,j)$ for $n = 1, \dots, N$ and $j = 1, \dots, J$.
The hyperparameters $\alpha$ and $\beta$ are scalar coefficients, $f_{\text{act}}$ denotes the activation function, and $\odot$ denotes the Hadamard product.
The features at the last layer $\mbf_{nj}^{(L)}$ are enforced to be in $\mathbb{R}^2$ such that the elements correspond to the real and imaginary parts of the entry $v^\prime_{nj}$ in the $n$-th row and $j$-th column of the matrix containing the non-normalized precoders $\mbV^\prime = [\mbv^\prime_1, \dots, \mbv^\prime_J]$, i.e., $\mbf_{nj}^{(L)} = [\Re(v^\prime_{nj}), \Im(v^\prime_{nj})]^\tp$. 
After normalizing the matrix $\mbV^\prime$ to satisfy the power constraint, the precoders $\{\mbv_j\}_{j=1}^J$ are obtained.
In \Cref{fig:stat_gnn_framework}, the block diagram of the proposed \ac{GNN}-based precoder design approach utilizing statistical knowledge is depicted.

During the training stage, the sum-rate expression from \eqref{eq:inst_sumrate} is used as the negative loss function, and we utilize the training set $\mathcal{D}_\text{genie} = \{ \{(\mbc_j^{(d)}, \mbh_j^{(d)}) \}_{j=1}^J  \}_{d=1}^D$, i.e., the training set consists of in total $D$ multi-user scenarios of $J$ \acp{MT}, where for each \ac{MT} the perfect statistical knowledge $\mbC_{\mbdelta_j}^{(d)}$ and the corresponding channel $\mbh_j^{(d)}$ is available.
The \ac{GNN} generalizes to different numbers of \acp{MT} by construction.
To enable operation at different \ac{SNR} levels, we train the \ac{GNN} for an \ac{SNR} range from $\SI{0}{dB}$ to $\SI{20}{dB}$ by adjusting the noise variance accordingly.
Thus, only one \ac{GNN} model is required to design the precoders for a varying number of \acp{MT} and different \ac{SNR} levels, which is beneficial in terms of memory.
Since the \ac{GNN} consists of a few layers only, and the feature updates in each layer can be processed in parallel, the inference time can be accelerated, where the computational complexity of the \ac{GNN} module is then $\mathcal{O}(\sum_{\ell=1}^{L} M_\ell M_{\ell-1})$ \cite{RiBeJoUt24}.

However, assuming perfect statistical knowledge is not practical. 
In the following, we adapt the \ac{GNN}-based precoder design framework to systems with approximate statistical knowledge obtained via the generative modeling-based feedback scheme in \cite{TuFeKoJoUt23}.
This scheme utilizes a \ac{GMM} as a generative prior, enabling the inference and feedback of approximate statistical information through limited feedback with $B$ bits.

\subsection{Combining GMMs and GNNs for Limited Feedback-based Precoder Design} \label{sec:comb_gnn_gmm}

Each channel $\mbh$ in the \ac{BS} environment follows an unknown \ac{PDF} $f_{\mbh}$. 
The channels $\mbh_j$ of \acp{MT} within the \ac{BS}' coverage area are thus treated as realizations of a random variable with \ac{PDF} $f_{\mbh}$, though this \ac{PDF} is not analytically available.
To approximate $f_{\mbh}$ given a set of representative channel samples, the method in \cite{TuFeKoJoUt23} employs a \ac{GMM} with $K=2^B$ components
\begin{equation}\label{eq:gmm_of_h}
    f^{(K)}_{\mbh}(\mbh_j) = \sum\nolimits_{k=1}^K \pi_k \calN_{\C}(\mbh_j; \mbmu_k, \mbC_k),
\end{equation}
encoding environmental knowledge through conditional Gaussian distributions.
The means of the \ac{GMM}-components are set to zero, i.e., $\mbmu_k=\bm{0}$ for all $k \in \{1,\dots, K\}$, where~\cite{BoBaTuSe24} provides a theoretical justification for this regularization.
Additionally, depending on the \ac{BS} antenna array geometry, the \ac{GMM} covariance matrices are constrained to the structure $\mbC_{k} = \mbQ^\herm \diag(\mbq_{k}) \mbQ$.
For a \ac{ULA} with $N$ elements, $\mbQ$ contains the first $N$ columns of a $2N\times 2N$ \ac{DFT} matrix and $\mbq_{k} \in \R_{+}^{2N}$, such that the \ac{GMM} covariance matrices are Toeplitz matrices.
For a \ac{URA} with $\Nv$ vertical and $\Nh$ horizontal ($N=\Nv\Nh$) elements, we have $\mbQ = \mbQ_{\Nv} \otimes \mbQ_{\Nh}$, where $\mbQ_T$ (with $T \in \{\Nv, \Nh\}$) contains the first $T$ columns of a $2T\times 2T$ \ac{DFT} matrix, and $\mbq_{k} \in \R_{+}^{4N}$, such that the \ac{GMM} covariance matrices are block-Toeplitz matrices with Toeplitz blocks \cite{FeJoHuKoTuUt22}.
The structured covariance matrices reduce the number of learnable parameters, mitigating overfitting.
The \ac{GMM} parameters, i.e., $\{\pi_k, \mbC_k\}_{k=1}^K$, are shared among all \acp{MT}, and are learned in an offline phase.
Maximum likelihood estimates of the \ac{GMM} parameters are obtained from a training set \(\mathcal{H} \), using an \ac{EM} algorithm with an adapted M-step following \cite{FeJoHuKoTuUt22} and the means set to zero.

In the online phase, each \ac{MT} infers its feedback information from the pilot observation $\mby_j$ using the \ac{GMM} of the observation
\begin{equation}\label{eq:gmm_y}
    f_{\mby}^{(K)}(\mby_j) = \sum\nolimits_{k=1}^K \pi_k \calN_{\C}(\mby_j; \mathbf{0}, \mbP \covhk \mbP^\herm + \mbSigma).
\end{equation}
Each \ac{MT} $j$ then computes the responsibilities for $\mby_j$ as follows
\begin{equation}\label{eq:responsibilities}
    p(k \mid \mby_j) = \frac{\pi_k \calN_{\C}(\mby_j; \mathbf{0}, \mbP \covhk \mbP^\herm + \mbSigma)}{\sum_{i=1}^K \pi_i \calN_{\C}(\mby_j; \mathbf{0}, \mbP \covhi \mbP^\herm + \mbSigma) },
\end{equation}
and selects the feedback index $k_j^\star$ through a \ac{MAP} estimation \cite{TuFeKoJoUt23}:
\begin{equation} \label{eq:ecsi_index_j}
    k^\star_j = \argmax_{k } ~{p(k \mid \mby_j)}
\end{equation}
Evaluating \eqref{eq:ecsi_index_j} does not scale with $N$ and has a complexity of only \( \calO(K n_{\mathrm{p}}^2) \), a key advantage in massive \ac{MIMO} systems~\cite{TuFeKoJoUt23}.

For performance reference, perfect \ac{CSI} can be assumed at each \ac{MT} to determine the feedback index as \cite{TuFeKoJoUt23}:
\begin{equation} \label{eq:pcsi_index}
    k^\star_j = \argmax_{k } ~{p(k \mid \mbh_j)}
\end{equation}

After the \acp{MT} send their feedback indices to the \ac{BS}, the \ac{GMM}-based statistical information for each \ac{MT}, i.e., assuming $\mbh_j$ is distributed as $\calN_{\C}(\mathbf{0}, \mbC_{k^\star_j})$, is utilized for precoder design via the \ac{GNN}.
Due to the structured \ac{GMM} covariance matrices, each $\mbC_k$ is fully represented by its first row.
Thus, for each \ac{MT}, the first row of the respective \ac{GMM} component's covariance matrix, $\mbc_{j} = \mbC_{k^\star_j}[1,:]$, is extracted, of which the real and imaginary parts are stacked, and passed through the feature extractor $\mbF_\theta$ to obtain the input feature vectors for the \ac{GNN}.
The \ac{GNN} output then provides the precoders $\{\mbv_j\}_{j=1}^J$ after normalization, cf. \Cref{fig:stat_gnn_framework} and \Cref{sec:gnn_perfect}.

To account for the \ac{GMM}-based feedback for precoder design using the \ac{GNN}, the training set $\mathcal{D} = \{ \{(\mbc_j^{(d)}, \mbh_j^{(d)}) \}_{j=1}^J  \}_{d=1}^D$ again includes $D$ multi-user scenarios with $J$ \acp{MT}.
For each \ac{MT}, statistical information $\mbc_j^{(d)} = \mbC_{k^\star_j}^{(d)}[1,:]$ is obtained through a pilot observation $\mby_j^{(d)}$ using the \ac{GMM} via \eqref{eq:ecsi_index_j}, and the corresponding channel $\mbh_j^{(d)}$ is available in the offline training phase. 
When assuming perfect \ac{CSI} at each \ac{MT}, $\mbc_j^{(d)}$ is obtained utilizing \eqref{eq:pcsi_index}.
The \ac{GNN} is trained for an \ac{SNR} range from $\SI{0}{dB}$ to $\SI{20}{dB}$ and enables precoder design for a varying number of \acp{MT}.
Moreover, the \ac{GMM}-based feedback scheme allows multi-user operation for a varying number of \acp{MT} $J$, and due to the analytic representation of the \ac{GMM}, facilitates adaptation to any \ac{SNR} level, without requiring re-training.
This combination of the \ac{GMM}-based feedback scheme and the \ac{GNN}-based precoder design yields a fully flexible approach for multi-user systems, where the \ac{GMM} knowledge is required at both the \acp{MT} and the \ac{BS}, while the \ac{GNN} knowledge is only required at the \ac{BS}.
An overview of the combined \ac{GMM}- and \ac{GNN}-based precoder design scheme is provided in \Cref{alg:gmm_gnn}.

\begin{algorithm}[t]
\captionsetup{font=footnotesize}
\caption{Precoder Design via Graph Neural Network using Gaussian Mixture Model-based Feedback Scheme}
\label{alg:gmm_gnn}
    \begin{algorithmic}[1] \footnotesize
    \REQUIRE Offline trained GMM and GNN, where the GMM is available at both the \acp{MT} and the \ac{BS}, and the GNN at the \ac{BS}.
    
    \hspace{-0.6cm}{\textbf{GMM-based Feedback Inference at the MTs}}
    
    \STATE Infer feedback index $k_j^\star$ from observation $\mby_j$ at each MT $j$
    
    $k^\star_j = \argmax_{k } ~{p(k \mid \mby_j)} \ \forall j$
    
    \STATE Each MT $j$ sends back $k^\star_j$ encoded by $B=2^K$ bits
    
    \color{black} \hspace{-0.6cm}{\textbf{GNN-based Precoder Design at the BS}}

    \STATE Extract the first row of the respective GMM component's covariance matrix for each MT $j$

    $\mbc_{j} = \mbC_{k^\star_j}[1,:] \ \forall j$

    \STATE Pass $\mbc_{k^\star_j}$ through feature extractor $\mbF_\theta$ for each MT $j$

    $\mbf_j^{(0)} = \mbF_\theta(\mbc_{k^\star_j}) \ \forall j$

    \STATE Provide feature vector of each MT $j$ to the GNN and obtain precoders
    
    $\{\mbv_j\}_{j=1}^J \gets \text{GNN}( \{\ \mbf_j^{(0)} \}_{j=1}^J )$
    \end{algorithmic}
\end{algorithm}

\section{Baseline Schemes for Precoder Design} \label{sec:baselines_swmmse_dft}

As a baseline, the \ac{SWMMSE} algorithm from \cite{RaSaLu16} is used for precoder design.
The \ac{SWMMSE} aims to maximize the ergodic sum-rate by treating the channels of all \acp{MT} as random variables and iteratively updates the precoders.
Each iteration $i$ requires a sample of every \ac{MT}'s channel for the update.
It was shown that the \ac{SWMMSE} converges to a set of stationary points of the ergodic sum-rate maximization problem for $i \to \infty$ \cite{RaSaLu16}.
For practical implementation, however, a finite maximum number of iterations, $I_{\max}$ is specified.
With perfect statistical knowledge, samples are generated as  \( \tilde\mbh_{j}^{(i)} \sim \calN_\C(\mbzero, \mbC_{\mbdelta_j}) \) for each \ac{MT} $j$ and provided to the \ac{SWMMSE}.
Alternatively, when statistical information for each \ac{MT} is obtained using the \ac{GMM} via \eqref{eq:ecsi_index_j} or \eqref{eq:pcsi_index}, samples are drawn as $\tilde\mbh_{j}^{(i)} \sim \calN_{\C}(\mbzero, \mbC_{k^\star_j})$ and fed to the \ac{SWMMSE}~\cite{TuFeKoJoUt23}.

Another baseline is the \ac{DFT} codebook-based limited feedback scheme, where each \ac{MT} estimates its channel $\hat{\mbh}_j$ and selects its feedback as $k^\star_j = \argmax_{k } |\mbw_k^\herm \hat{\mbh}_j|$ with $\mbw_k \in \mathcal{W}$, using a codebook $\mathcal{W} = \{\mbw_1, \dots, \mbw_{K} \}$ of size $|\mathcal{W}| = K = 2^B$~\cite{Ji06}.
Depending on whether a \ac{ULA} or \ac{URA} is deployed, a \ac{DFT} codebook or a 2D-\ac{DFT} codebook is used, respectively \cite{LiSuZeZhYuXiXu13}.
At the \ac{BS}, each channel is represented as $\hat{\mbh}^{(q)}_j = \mbw_{k^\star_j}$ \cite{Ji06}, and the \ac{IWMMSE} \cite[Algorithm~1]{HuCaYuQiXuYuDi21} is used.

\section{Simulation Results} \label{sec:sim_results}

In the case of the spatial channel model (see \Cref{sec:scm_data}), we simulate a \ac{BS} with $N=64$ antenna elements, assuming a single propagation cluster composed of multiple sub-paths per \ac{MT}.
To train the \ac{GNN} using perfect statistical knowledge, we generate the set $\mathcal{D}_{\text{genie}}$ with $D=2{,}400$ scenarios and $J=16$ \acp{MT}.
For training the \ac{GMM}, we create a set $\mathcal{H} = \{ \mbh^{(m)} \}_{m=1}^{M}$ with $M=10^5$.
For training the \ac{GNN} with \ac{GMM}-based feedback, we construct the set $\mathcal{D}$ with $D=2{,}400$ scenarios and $J=16$ \acp{MT} as described in \Cref{sec:comb_gnn_gmm}.
Note that the \acp{GNN} are solely trained for this configuration ($J=16$ \acp{MT})
and effectively generalize to configurations with $J<16$ \acp{MT}.
The \ac{GNN} consists of five hidden layers with dimension $M_\ell = 128$ for $\ell=1,\dots,L-1$, and we set $\alpha=0.1/N$, $\beta=0.1$, and use the ReLU activation function, similar to \cite{RiBeJoUt24}. 
We train the \ac{GNN} for $500$ epochs, where the batch size is $100$, and use the Adam optimizer with $0.001$ as the initial learning rate.
For model selection, a validation set with $D_\text{val}=300$ scenarios is used to choose the best-performing \ac{GNN} configuration.
To evaluate performance, we use the average sum-rate metric across similarly generated test sets, each with ${D}_{\text{test}}=500$ scenarios, for various numbers of \acp{MT} $J$.
When using measurement data (see \Cref{sec:meas_data}), we use the same data set sizes for training and validating the \ac{GNN}, for training the \ac{GMM}, and for testing.
For the spatial channel model and the measurement data, the channels are normalized to satisfy \( \expec[\|\mbh\|^2] = N \).
Additionally, we set $\rho=1$, enabling the \ac{SNR} to be defined as \( \frac{1}{\sigma_n^2} \).

For clarity, we omit the index $j$ in the legend descriptions below.
In the following discussion, ``\{GNN, SWMMSE\}, genie'', refers to the baselines where perfect statistical knowledge is assumed to design the precoders, see \Cref{sec:gnn_perfect} and \Cref{sec:baselines_swmmse_dft}, respectively.
The labels ``GNN, GMM, \{$\mbh, \mby$\},'' represent the cases where either perfect \ac{CSI}, see~\eqref{eq:pcsi_index}, is assumed at each \ac{MT}, or the observation $\mby_j$ at each \ac{MT} is used, see~\eqref{eq:ecsi_index_j}, for determining a feedback index via the \ac{GMM}-based feedback scheme and subsequently, the precoders are designed using the corresponding \ac{GNN}.
Similarly, ``SWMMSE, GMM, \{$\mbh, \mby$\},'' indicates that the \ac{SWMMSE} is used instead of the \ac{GNN} to design the precoders based on the \ac{GMM} feedback.
``IWMMSE, DFT, \{$\hhat_{\text{GMM}}$, $\hhat_{\text{LS}}$\},'' refers to the cases where the channel is estimated at each \ac{MT}, afterward, the feedback information is determined using the DFT codebook, see~\Cref{sec:baselines_swmmse_dft}.
We either use the \ac{LS} estimator $\hhat_{\text{LS}}$, or the \ac{GMM}-based estimator \( \hhat_{\text{GMM}} \) from \cite{KoFeTuUt21J}.
For both the \ac{SWMMSE} and the \ac{IWMMSE}, the maximum number of iterations is set to $I_{\max}=300$, unless otherwise specified.

\begin{figure}[tb]
    \centering
    \includegraphics[]{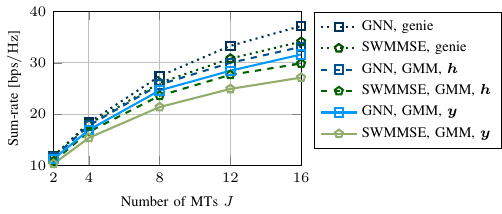}
    %\vspace{-0.25cm}
    \caption{The sum-rate over the number of \acp{MT} $J$ for a system with $B=6$ feedback bits, $n_\mathrm{p}=16$ pilots, and $\text{SNR}=\SI{10}{dB}$ (spatial channel model).}
    \label{fig:fig_3gpp_snr10_np16_overuser}
    \vspace{-0.5cm}
\end{figure}

\begin{figure}[tb]
    \centering
    \includegraphics[]{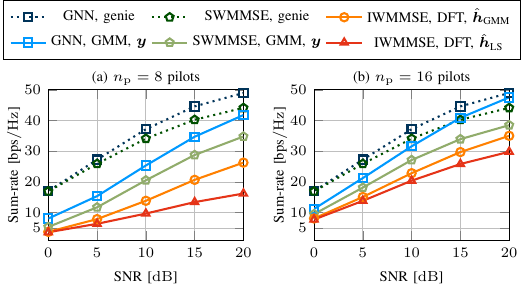}
    \vspace{-0.6cm}
    \caption{The sum-rate over the \ac{SNR} for a system with $B=6$ feedback bits, $J=16$ \acp{MT}, and (a) $n_\mathrm{p}=8$ pilots, or (b) $n_\mathrm{p}=16$ pilots. (spatial channel model).}
    \label{fig:fig_3gpp_np8_np16_J16_oversnr_comb}
    \vspace{-0.5cm}
\end{figure}

In \Cref{fig:fig_3gpp_snr10_np16_overuser}, \Cref{fig:fig_3gpp_np8_np16_J16_oversnr_comb}(a), and \Cref{fig:fig_3gpp_np8_np16_J16_oversnr_comb}(b), we use the spatial channel model from \Cref{sec:scm_data}.
In \Cref{fig:fig_3gpp_snr10_np16_overuser}, we fix $B=6$ bits, $n_\mathrm{p}=16$ pilots, and $\text{SNR}=\SI{10}{dB}$ and vary the number $J$ of served \acp{MT}.
We can observe that the proposed \ac{GNN}-based precoder design method using perfect statistical knowledge ``GNN, genie'' outperforms the baseline ``SWMMSE, genie'' for all considered numbers $J$ of served \acp{MT}.
The method ``GNN, GMM, $\mbh$'' shows performance degradation due to quantized feedback information associated with \ac{GMM} components, a trend also seen in the ``SWMMSE, GMM, $\mbh$'' baseline.
For the respective counterparts solely utilizing the observations for feedback inference, the proposed ``GNN, GMM, $\mby$,'' significantly outperforms the baseline ``SWMMSE, GMM, $\mby$.''
While the \ac{SWMMSE} relies on imperfect statistical input from the \ac{GMM} feedback for optimization, the \ac{GNN} benefits from access to \ac{CSI} during the offline training. 
This allows us to accurately evaluate the loss function of the \ac{GNN} and perform updates, a capability that is not given by the \ac{SWMMSE}.
Consequently, the \ac{GNN} holds a significant advantage over the \ac{SWMMSE}, despite both using the same input during the online phase.

In \Cref{fig:fig_3gpp_np8_np16_J16_oversnr_comb}(a), we set $B=6$ bits, $J=16$ \acp{MT}, and $n_\mathrm{p}=8$ pilots and depict the sum-rate over the \ac{SNR}.
While at low \ac{SNR} values ``GNN, genie'' and ``SWMMSE, genie'' perform equally well, with an increasing \ac{SNR}, the proposed ``GNN, genie'' method is superior.
Moreover, the proposed ``GNN, GMM, $\mby$'' approach outperforms the baseline ``SWMMSE, GMM, $\mby$'' based on the same \ac{GMM} feedback and all other baselines ``IWMMSE, DFT, \{$\hhat_{\text{GMM}}$, $\hhat_{\text{LS}}$\}.''
While the \ac{GMM} estimator improves the DFT-based feedback scheme over the \ac{LS} estimator, the performance is still significantly worse than the ``GNN, GMM, $\mby$'' approach.
Remarkably, in \Cref{fig:fig_3gpp_np8_np16_J16_oversnr_comb}(b) where we have the same setup but with $n_\mathrm{p}=16$, the proposed ``GNN, GMM, $\mby$'' approach even outperforms the ``SWMMSE, genie'' in the high \ac{SNR} regime, due to more accurate \ac{GMM} feedback with the additional pilots, underlining the effectiveness of the proposed approach in low pilot overhead systems.

\begin{figure}[tb]
    \centering
    \includegraphics[]{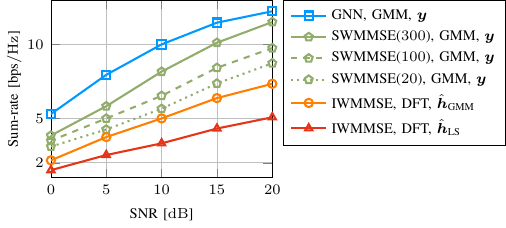}
    \vspace{-0.2cm}
    \caption{The sum-rate over the \ac{SNR} for a system with $B=6$ feedback bits, $J=8$ \acp{MT}, and $n_{\mathrm{p}}=8$ pilots (measured channel data).}
    \label{fig:fig_meas_np8_oversnr}
    \vspace{-0.5cm}
\end{figure}

Lastly, in \Cref{fig:fig_meas_np8_oversnr}, we use the measurement data (see \Cref{sec:meas_data}) to simulate a system with $B=6$ bits, $J=8$ \acp{MT}, and $n_\mathrm{p}=8$ pilots and depict the sum-rate over the \ac{SNR}.
In this case as well, we observe that the proposed ``GNN, GMM, $\mby$'' approach outperforms all baselines.
Since we do not have access to perfect statistics in the case of measured channels, the respective curves are missing.
To further emphasize the advantage of the parallelizable \ac{GNN}, where precoders are determined in a single forward pass, we also evaluate the computationally intensive iterative \ac{SWMMSE} with reduced maximum iteration counts of 20 and 100 (in addition to $I_{\max}=300$), denoted as ``SWMMSE($\{20, 100, 300\}$), GMM, $\mby$'' for faster precoder calculation.
We observe that with fewer iterations, the performance of the \ac{SWMMSE} baseline degrades.

\section{Conclusion \& Outlook}

In this letter, we proposed a \ac{GNN}-based framework for statistical precoder design, leveraging model-based insights to achieve a compact representation of statistical knowledge, resulting in lightweight architectures.
The framework is extended to handle approximate statistical information in \ac{FDD} systems via a \ac{GMM}-based limited feedback scheme, and demonstrates strong performance in simulations.
Future work could explore alternative generative models like \ac{VQ-VAE}-based scheme from \cite{TuBaLiUt24} and develop an end-to-end system that additionally learns the pilot matrix.

\bibliographystyle{IEEEtran}
\bibliography{IEEEabrv,biblio}
\end{document}